\documentclass[12pt,letterpaper]{article}
\usepackage{amsmath,amssymb,array,calc,rotating,epsfig,psfrag,amscd, cite}

\makeatletter
\renewcommand\section{\@startsection {section}{1}{\z@}%
                                   {-3.5ex \@plus -1ex \@minus -.2ex}
                                   {2.3ex \@plus.2ex}%
                                   {\normalfont\large\bfseries}}
\renewcommand\subsection{\@startsection{subsection}{2}{\z@}%
                                     {-3.25ex\@plus -1ex \@minus -.2ex}%
                                     {1.5ex \@plus .2ex}%
                                     {\normalfont\bfseries}}
\makeatother

\let\non\nonumber

\let\s=\sigma

\let\S=\Sigma

\newcommand{\bea}{\begin{eqnarray}}
\newcommand{\eea}{\end{eqnarray}}
\newcommand{\be}{\begin{equation}}
\newcommand{\ee}{\end{equation}}

\newcommand{\p}{\partial}

\newcommand{\C}[1]{$(\ref{#1})$}

\def\IZ{\relax\ifmmode\mathchoice
{\hbox{\cmss Z\kern-.4em Z}}{\hbox{\cmss Z\kern-.4em Z}}
{\lower.9pt\hbox{\cmsss Z\kern-.4em Z}} {\lower1.2pt\hbox{\cmsss
Z\kern-.4em Z}}\else{\cmss Z\kern-.4em Z}\fi}
\def\IR{\relax{\rm I\kern-.18em R}}

\def\one{{\hbox{ 1\kern-.8mm l}}}

\newlength{\bredde}
\def\slash#1{\settowidth{\bredde}{$#1$}\ifmmode\,\raisebox{.15ex}{/}
\hspace*{-\bredde} #1\else$\,\raisebox{.15ex}{/}\hspace*{-\bredde}
#1$\fi}

\newsavebox{\zzzbar}
\sbox{\zzzbar}
  {\setlength{\unitlength}{0.9em}
  \begin{picture}(0.6,0.7)
  \thinlines
  \put(0,0){\line(1,0){0.6}}
  \put(0,0.75){\line(1,0){0.575}}
  \multiput(0,0)(0.0125,0.025){30}{\rule{0.3pt}{0.3pt}}
  \multiput(0.2,0)(0.0125,0.025){30}{\rule{0.3pt}{0.3pt}}
  \put(0,0.75){\line(0,-1){0.15}}
  \put(0.015,0.75){\line(0,-1){0.1}}
  \put(0.03,0.75){\line(0,-1){0.075}}
  \put(0.045,0.75){\line(0,-1){0.05}}
  \put(0.05,0.75){\line(0,-1){0.025}}
  \put(0.6,0){\line(0,1){0.15}}
  \put(0.585,0){\line(0,1){0.1}}
  \put(0.57,0){\line(0,1){0.075}}
  \put(0.555,0){\line(0,1){0.05}}
  \put(0.55,0){\line(0,1){0.025}}
  \end{picture}}

\newcommand{\ena}{\end{eqnarray}}
\newcommand{\beqa}{\begin{eqnarray}}
\newcommand{\eeqa}{\end{eqnarray}}

\def\s{\sigma}

\def\S{\Sigma}

\begin{document}
\begin{titlepage}

\begin{center}



\vskip 2 cm
{\Large \bf Poisson equation for the Mercedes diagram in string theory at genus one}\\
\vskip 1.25 cm { Anirban Basu\footnote{email address:
    anirbanbasu@hri.res.in} } \\
{\vskip 0.5cm Harish--Chandra Research Institute, Chhatnag Road, Jhusi,\\
Allahabad 211019, India\\}

\end{center}

\vskip 2 cm

\begin{abstract}
\baselineskip=18pt

The Mercedes diagram has four trivalent vertices which are connected by six links such that they form the edges of a tetrahedron. This three loop Feynman diagram contributes to the $D^{12} \mathcal{R}^4$ amplitude at genus one in type II string theory, where the vertices are the points of insertion of the graviton vertex operators, and the links are the scalar propagators on the toroidal worldsheet. We obtain a modular invariant Poisson equation satisfied by the Mercedes diagram, where the source terms involve one and two loop Feynman diagrams. We calculate its contribution to the $D^{12} \mathcal{R}^4$ amplitude.

\end{abstract}

\end{titlepage}


\section{Introduction}

Perturbative amplitudes in superstring theory in a certain background contain invaluable information about string interactions. When expanded around weak string coupling for a certain compactification, this expansion is an asymptotic expansion which at genus $g$ is of the form $(e^{-2\phi} V)^{1-g} f(\lambda_i)$, where $\phi$ is the dilaton, $V$ is the volume of the internal manifold in the string frame metric, and $\lambda_i$ are the various other moduli of the compactification which show up in the perturbative amplitudes. Along with non--perturbative corrections, these amplitudes yield exact S--matrix elements leading to various interactions in the low energy effective action. Though these terms in the effective action are difficult to determine in general, they can be determined in some cases exactly. These include BPS interactions in toroidally compactified type II string theory which preserve maximal supersymmetry~\cite{Green:1997tv,Green:1997as,Kiritsis:1997em,Green:1998by,Green:1999pu,D'Hoker:2005jc,D'Hoker:2005ht,Matone:2005vm,Berkovits:2005ng,Green:2005ba,Berkovits:2006vc,Basu:2007ru,Basu:2007ck,Green:2008bf,Basu:2008cf,Green:2010kv,Green:2010wi,Basu:2011he,D'Hoker:2013eea,Gomez:2013sla,D'Hoker:2014gfa,Basu:2014hsa,Bossard:2014aea,Pioline:2015yea,Bossard:2015uga,Bossard:2015oxa,Basu:2015dqa,Pioline:2015nfa,Bossard:2015foa}. In the Einstein frame, this leads to U--duality covariant equations of motion. Hence, these perturbative amplitudes are useful not only to evaluate perturbative parts of the various S--matrices, but also to understand the role of the non--perturbative U--duality symmetries of the theory. 

Let us consider the perturbative amplitude at genus one in type II string theory in ten dimensional flat spacetime, where the external states involve four on--shell graviton vertex operators. This amplitude is the same in the type IIA and type IIB theories. This has analytic as well as non--analytic terms in the $\alpha'$ expansion. The total amplitude is given by an integral over the fundamental domain of $SL(2,\mathbb{Z})$. While the analytic terms have polynomial dependence on the Mandelstam variables, the non--analytic ones have logarithmic dependence on them. The analytic contribution at every order in the $\alpha'$ expansion involves an integral of the form~\cite{Green:1999pv,Green:2008uj,D'Hoker:2015foa}
\be \int_{\mathcal{F}_L} \frac{d^2\tau}{\tau_2^2} f(\tau,\bar\tau) ,\ee         
where
\be \mathcal{F}_L = \{ -\frac{1}{2} \leq \tau_1 \leq \frac{1}{2}, \quad \vert \tau_2 \vert \leq L\},\ee
and one takes $L\rightarrow \infty$. On the other hand, the non--analytic contribution involves an integral over $\mathcal{R}_L$ defined by
\be \mathcal{R}_L = \{ -\frac{1}{2} \leq \tau_1 \leq \frac{1}{2}, \quad \vert \tau_2 \vert > L\},\ee
with appropriate integrands depending on the amplitude. Note that $\mathcal{F}_L \oplus \mathcal{R}_L$ is the fundamental domain of $SL(2,\mathbb{Z})$.

Thus the analytic contributions are obtained by integrating over the truncated fundamental domain of $SL(2,\mathbb{Z})$, and the integral yields terms finite as well as divergent in this limit. The non--analytic contributions are obtained from the boundary of moduli space as $L\rightarrow \infty$ in the integral over $\mathcal{R}_L$. Both the integrals over $\mathcal{F}_L$ and $\mathcal{R}_L$ have terms that diverge as $L\rightarrow \infty$, but the total divergence cancels at each order in the $\alpha'$ expansion. The remaining finite contributions are the contributions to the one loop amplitude.  

We shall be concerned with analytic terms obtained by integrating over $\mathcal{F}_L$ for the four graviton amplitude. They involve modular invariant integrands which satisfy Poisson equations. These integrands are completely determined by the topology of the Feynman diagrams resulting from joining the positions of the vertex operators by scalar propagators on the toroidal worldsheet in various ways. Using these Poisson equations, the contribution of the four graviton amplitude upto the $D^{10}\mathcal{R}^4$ interaction in the low energy expansion has been worked out~\cite{D'Hoker:2015foa,D'Hoker:2015zfa}. Among the several worldsheet Feynman diagrams that contribute to the $D^{12}\mathcal{R}^4$ interaction, we consider the contribution of the Mercedes diagram to the ten dimensional amplitude. Note that these amplitudes, beyond the $D^6\mathcal{R}^4$ interaction, contribute to non--BPS terms in the effective action. 

We begin with a brief review of the genus one four graviton amplitude in the type II theory in ten dimensions. This is followed by a general discussion of the moduli dependence of the integrand, and how we propose to analyze it using Beltrami differentials. We then derive the Poisson equation satisfied by the Mercedes diagram, which involves four vertices and six propagators connecting them, such that every vertex is trivalent. This is a three loop Feynman diagram. We show that the Poisson equation for this diagram involves source terms with one and two loop Feynman diagrams. Finally, we calculate its contribution to the $D^{12}\mathcal{R}^4$ amplitude by integrating over $\mathcal{F}_L$ and keeping the finite terms as $L\rightarrow \infty$.

\section{The general structure of the type II one loop four graviton amplitude}

The one loop four graviton amplitude in type II superstring theory is given by
\be \mathcal{A}_4^{(1)} = 2\pi \mathcal{I}(s,t,u) \mathcal{R}^4, \ee
where
\be \label{oneloop}\mathcal{I} (s,t,u) = \int_{\mathcal{F}} \frac{d^2\tau}{\tau_2^2} F(s,t,u;\tau,\bar\tau),\ee
where we have integrated over $\mathcal{F}$, the fundamental domain of $SL(2,\mathbb{Z})$. The Mandelstam variables $s, t, u$ satisfy the on--shell condition
\be s+t+u=0.\ee 
We have defined $d^2 \tau = d\tau_1 d\tau_2$. The factor $F(s,t,u;\tau,\bar\tau)$ which encodes the moduli dependence is given by
\be \label{D}F(s,t,u;\tau,\bar\tau) = \prod_{i=1}^4 \int_\S  \frac{d^2 z^{(i)}}{\tau_2} e^{\mathcal{D}}.\ee
Here $z^{(i)}$ $(i=1,2,3,4)$ are the positions of insertions of the four vertex operators on the toroidal worldsheet $\S$. Hence $d^2 z^{(i)} = d({\rm Re} z^{(i)}) d({\rm Im}z^{(i)})$, where
\be -\frac{1}{2} \leq {\rm Re} z^{(i)} \leq \frac{1}{2}, \quad 0 \leq {\rm Im} z^{(i)}\leq \tau_2 \ee
for all $i$. In \C{D}, the expression for $\mathcal{D}$ is given by
\be \label{defD}4\mathcal{D} = \alpha' s (\hat{G}_{12} + \hat{G}_{34})+\alpha' t (\hat{G}_{14} + \hat{G}_{23})+ \alpha' u (\hat{G}_{13} +\hat{G}_{24}),\ee
where $\hat{G}_{ij}$ is the scalar Green function on the torus with complex structure $\tau$ between points $z^{(i)}$ and $z^{(j)}$, and so
\be \hat{G}_{ij} \equiv \hat{G}(z^{(i)} - z^{(j)};\tau).\ee
In particular, it is defined as~\cite{Green:1999pv,Green:2008uj}
\bea \label{prop}\hat{G}(z;\tau) &=& -{\rm ln} \Big\vert \frac{\theta_1 (z\vert\tau)}{\theta_1'(0\vert\tau)} \Big\vert^2 + \frac{2\pi({\rm Im}z)^2}{\tau_2} \non \\ &=& \frac{1}{\pi} \sum_{(m,n)\neq(0,0)} \frac{\tau_2}{\vert m\tau+n\vert^2} e^{\pi[\bar{z}(m\tau+n)-z(m\bar\tau+n)]/\tau_2} + 2{\rm ln} \vert \sqrt{2\pi} \eta(\tau)\vert^2.\eea
Now the $z$ independent zero mode part given by the second term in the second line of \C{prop} cancels in the whole amplitude, which follows from the expression for $\mathcal{D}$ in \C{defD} on using $s+t+u=0$. Thus in the expression for $\mathcal{D}$ we simply replace $\hat{G}(z;\tau)$ by $G(z;\tau)$ where
\be \label{Green}G(z;\tau) = \frac{1}{\pi} \sum_{(m,n)\neq(0,0)} \frac{\tau_2}{\vert m\tau+n\vert^2} e^{\pi[\bar{z}(m\tau+n)-z(m\bar\tau+n)]/\tau_2}.\ee
Note that $G(z;\tau)$ is modular invariant, and single valued. Thus
\be \label{sv}G(z;\tau) = G(z+1;\tau) = G(z+\tau;\tau).\ee
As explained before, in \C{oneloop}, $\mathcal{F}$ is split into
\be \mathcal{F} = \mathcal{F}_L +\mathcal{R}_L,\ee
where $\mathcal{F}_L$ is defined for $\tau_2 \leq L$, and $\mathcal{R}_L$ is defined for $\tau_2 > L$. Thus the analytic part of the amplitude is given by
\be \mathcal{I}_{an} (s,t,u) =   \sum_{n=0}^\infty\int_{\mathcal{F}_L} \frac{d^2\tau}{\tau_2^2}\prod_{i=1}^4 \int_\S  \frac{d^2 z^{(i)}}{\tau_2} \cdot \frac{\mathcal{D}^n}{n!}.\ee
Hence performing an $\alpha'$ expansion we get that
\be \mathcal{I}_{an} (s,t,u)= \sum_{p,q} \s_2^p \s_3^q J^{p,q},\ee
where
\be J^{p,q} = \int_{\mathcal{F}_L} \frac{d^2\tau}{\tau_2^2} j^{p,q} (\tau,\bar\tau),\ee
and
\be \s_2 = \alpha'^2 (s^2 +t^2 +u^2), \quad \s_3 = \alpha'^3 (s^3 +t^3 +u^3).\ee
Here $j^{p,q}(\tau,\bar\tau)$ is obtained after integrating over the insertion points of the vertex operators and encodes the topologically distinct ways the scalar propagators are connected on the toroidal worldsheet. These contributions upto the $D^{10}\mathcal{R}^4$ term in the low energy expansion have been considered in detail~\cite{D'Hoker:2015foa,D'Hoker:2015zfa}. The integrands satisfy Poisson equations with specific source terms. The structure of several of these equations was conjectured in~\cite{D'Hoker:2015foa}, while one of which was proven in~\cite{D'Hoker:2015zfa} using elaborate calculations. We shall consider a particularly simple diagram at $O(D^{12}\mathcal{R}^4)$ in the derivative expansion, for which we derive the Poisson equation it satisfies. This is given by the Mercedes diagram in figure 1. This is a three loop Feynman diagram where the six scalar propagators connect all the four graviton vertices such that each vertex is trivalent. Note that the six links of the diagram form the edges of a tetrahedron. Thus we have that
\be \label{M}\mathcal{M} = \int_\S  \frac{d^2 z^{(1)}}{\tau_2} \int_\S  \frac{d^2 z^{(2)}}{\tau_2} \int_\S  \frac{d^2 z^{(3)}}{\tau_2}  \int_\S \frac{d^2 z^{(4)}}{\tau_2} G_{12} G_{13} G_{14} G_{23} G_{34} G_{14}. \ee

\begin{figure}[ht]
\begin{center}
\[
\mbox{\begin{picture}(120,90)(0,0)
\includegraphics[scale=.45]{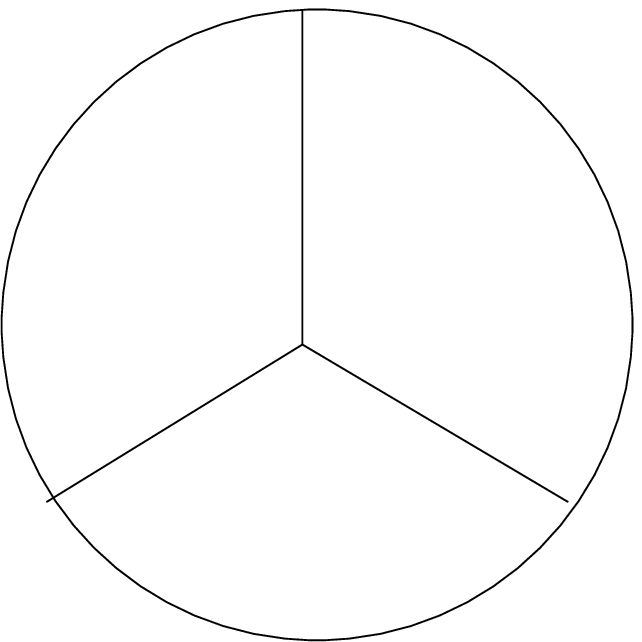}
\end{picture}}
\]
\caption{The Mercedes diagram $\mathcal{M}$}
\end{center}
\end{figure}

Its contribution to the $D^{12}\mathcal{R}^4$ interaction is given by~\cite{Green:2008uj}
\bea \label{cont}j^{(0,2)} = \frac{80}{6!}\mathcal{M} +\ldots.\eea    

In order to find the Poisson equation satisfied by $\mathcal{M}$, we find it useful to define the non--holomorphic Eisenstein series
\be E_s (\tau,\bar\tau) = \sum_{(m,n)\neq (0,0)} \frac{\tau_2^s}{\pi^s \vert m+n\tau\vert^{2s}}\ee
which satisfies the Laplace equation
\be \label{eisenstein}\Delta E_s (\tau,\bar\tau) = s(s-1) E_s (\tau,\bar\tau).\ee
Here the $SL(2,\mathbb{Z})$ invariant Laplacian is defined by
\be \Delta = 4\tau_2^2\frac{\p^2}{\p\tau\p\bar\tau}.\ee

\section{The systematics of the moduli dependence}

In order to obtain the differential equation satisfied by $\mathcal{M}$, we shall use the relations satisfied by the Green function $G_{ij}$ in \C{Green} under variations of the Beltrami differentials. To obtain them, we first write down these relations satisfied by $\hat{G}_{ij}$ at arbitrary genus and specialize to the case of genus one. Then on using \C{prop} this gives us the required relations for $G_{ij}$. 

In general, the scalar Green function on the genus $g$ Riemann surface is given by (see~\cite{D'Hoker:1988ta} for a detailed discussion)
\be \hat{G} (z,w) = -{\rm ln} \vert E(z,w)\vert^2 +2\pi Y^{-1}_{IJ} \Big({\rm Im}\int_z^w \omega_I \Big)\Big({\rm Im} \int_z^w \omega_J\Big).\ee
Here $E(z,w)$ is the prime form, $\omega_I$ is the Abelian differential one form, and the period matrix $\Omega$ is defined as $\Omega_{IJ} = X_{IJ} +iY_{IJ}$, where $X$ and $Y$ are matrices with real entries, and $I,J=1,\ldots,g$. We also have that $(Y^{-1})_{IJ} \equiv Y^{-1}_{IJ}$.

Now for the Beltrami differential $\mu$, the holomorphic variation $\delta_\mu \Psi$ for any $\Psi$ is given by
\be \label{Beltrami}\delta_\mu \Psi = \frac{1}{\pi} \int_\S d^2 z \mu_{\bar{z}}^{~z} \delta_{zz} \Psi.\ee 
Using the general expressions for the variation~\cite{Verlinde:1986kw,D'Hoker:1988ta}
\bea \delta_{ww} \omega_I (z) &=& \omega_I (w) \p_w\p_z {\rm ln} E(z,w), \non \\ \delta_{ww} \Omega_{IJ} &=& 2\pi i \omega_I (w) \omega_J (w), \non \\ \delta_{ww} {\rm ln} E(z_1,z_2) &=& -\frac{1}{2} \Big(\p_w {\rm ln}E(w,z_1)- \p_w {\rm ln}E(w,z_2)\Big)^2, \non \\ \delta_{\bar{w}\bar{w}} \p_z &=& \pi \delta^2 (z-w) \bar{\p}_z,\eea
we get that~\cite{D'Hoker:2014gfa}
\bea \label{defvar}\delta_{ww} \hat{G}(z_1,z_2) &=& \frac{1}{2} \Big(\p_w \hat{G}(w,z_1)- \p_w \hat{G}(w,z_2)\Big)^2, \non \\ \delta_{\bar{u}\bar{u}} \Big(\p_z \hat{G}(z,w)\Big) &=& -\frac{\pi}{2} \bar{\p}_z \delta^2 (z-u) +\pi Y^{-1}_{IJ} \omega_I (z) \overline{\omega_J (u)} \Big(\bar\p_u \hat{G} (u,w) -\bar\p_u \hat{G}(u,z)\Big).\non \\\eea

Now on the torus, the Beltrami differential $\mu_{\bar{z}}^{~z}$ is unity, $\omega(z) =dz$, and $\Omega= \tau$. 
The Green function satisfies
\bea \label{eigen}\bar{\p}_w\p_z G(z,w) = \pi \delta^2 (z-w) - \frac{\pi}{\tau_2}, \non \\ \bar{\p}_z\p_z G(z,w) = -\pi \delta^2 (z-w) + \frac{\pi}{\tau_2}. \eea
Now consider \C{Beltrami}  when $\Psi = \hat{G}(z_1,z_2)$. From \C{prop} we get that
\be \delta_\mu G(z_1,z_2) +4i\tau_2 \frac{\p{\rm ln}\eta(\tau)}{\p\tau} = \frac{1}{2\pi} \int_\S d^2 z \Big(\p_z G(z,z_1) - \p_z G(z,z_2)\Big)^2,\ee
where we have used
\be \int_\S d^2 w \delta_{ww} {\rm ln} \eta(\tau) = 2\pi i \tau_2 \frac{\p{\rm ln}\eta(\tau)}{\p\tau}.\ee
We also see that
\be \int_\S d^2 z \Big(\p_z G(z,w)\Big)^2 = -\tau_2 G_2 (\tau),\ee
where $G_2 (\tau)$ is the holomorphic Eisenstein series defined by
\be G_2 (\tau) = \sum_{(m,n) \neq (0,0)} \frac{1}{(m\tau+n)^2}.\ee
This leads to the simple relation
\be \label{onevar} \delta_\mu G(z_1,z_2) = -\frac{1}{\pi} \int_\S d^2 z \p_z G(z,z_1) \p_z G(z,z_2),\ee
on using the identity
\be \frac{\p{\rm ln}\eta(\tau)}{\p\tau} = \frac{iG_2 (\tau)}{4\pi}.\ee
Also from \C{onevar} we get that
\bea \delta_{\bar\mu} \delta_\mu G(z_1,z_2) &=& -\frac{1}{2\pi} \int_\S d^2 z \int_\S d^2 u \bar\p_u \delta^2 (z-u)  \p_z G(z,z_1) \non \\ &&- \frac{1}{\pi\tau_2} \int_\S d^2 z \int_\S d^2 u \p_z G(z,z_1)\bar\p_u \Big(G(u,z_2)- G(u,z)\Big)+ 1\leftrightarrow 2 \non \\ \eea
on using the second equation in \C{defvar}. The contribution from the second line vanishes as it is a total derivative in $\bar{u}$ and $G(u,z)$ is single valued. The contribution from the first line vanishes as well as it is a total derivative in $\bar{u}$ again, and the boundary contributions cancel by relabelling $z$ by $z+1$ and $z+\tau$ in the delta function and using the single valuedness of $G(z,z_1)$. Thus we get that
\be \label{twovar} \delta_{\bar\mu} \delta_\mu G(z_1,z_2)=0.\ee   
We shall use \C{onevar} and \C{twovar} repeatedly in our analysis below.
Finally the $SL(2,\mathbb{Z})$ invariant Laplacian is given by
\be \label{beltrami}\Delta = \delta_\mu \delta_{\bar\mu}.\ee

\section{The Poisson equation satisfied by the Mercedes diagram}

We now obtain the Poisson equation satisfied by $\mathcal{M}$ given in \C{M} using the analysis of the section above. In the various manipulations that are needed, we often obtain expressions involving $\p_z G(z,w)$ where $z$ is integrated over $\S$. We then integrate by parts without picking up boundary contributions on $\S$ as $G(z,w)$ is single valued. Also we readily use $\p_z G(z,w) = -\p_w G(z,w)$ using the translational invariance of the Green function. Finally, we have that
\be \int_\S d^2 z G(z,w)=0\ee
which follows from \C{Green}.

In the analysis below, for brevity we write
\be \int_\S d^2 z \int_\S d^2 w \ldots \equiv \int_{zw\ldots}.\ee

\subsection{Deriving the Poisson equation}

From \C{beltrami} we have that ($\mathcal{M}$ is referred to as $D_{1,1,1;1,1,1}$ in~\cite{Green:2008uj}.)
\be \Delta \mathcal{M} = \delta_\mu \delta_{\bar\mu} \mathcal{M} = 24 \mathcal{M}_1 + 6 \mathcal{M}_2,\ee 
where $\mathcal{M}_1$ and $\mathcal{M}_2$ are defined by
\bea \label{M12} \mathcal{M}_1 = \frac{1}{\tau_2^4} \int_{1234} \delta_\mu G_{12} \delta_{\bar\mu} G_{13} G_{14} G_{23} G_{34} G_{42}, \non \\ \mathcal{M}_2 = \frac{1}{\tau_2^4} \int_{1234} \delta_{\mu} G_{12} \delta_{\bar\mu} G_{34} G_{13} G_{14} G_{23} G_{24}.\eea

These two topologically distinct contributions are given in figure 2. Here $\mu$ along a link stands for $\delta_\mu$, while $\bar\mu$ stands for $\delta_{\bar\mu}$. 
\begin{figure}[ht]
\begin{center}
\[
\mbox{\begin{picture}(275,110)(0,0)
\includegraphics[scale=.55]{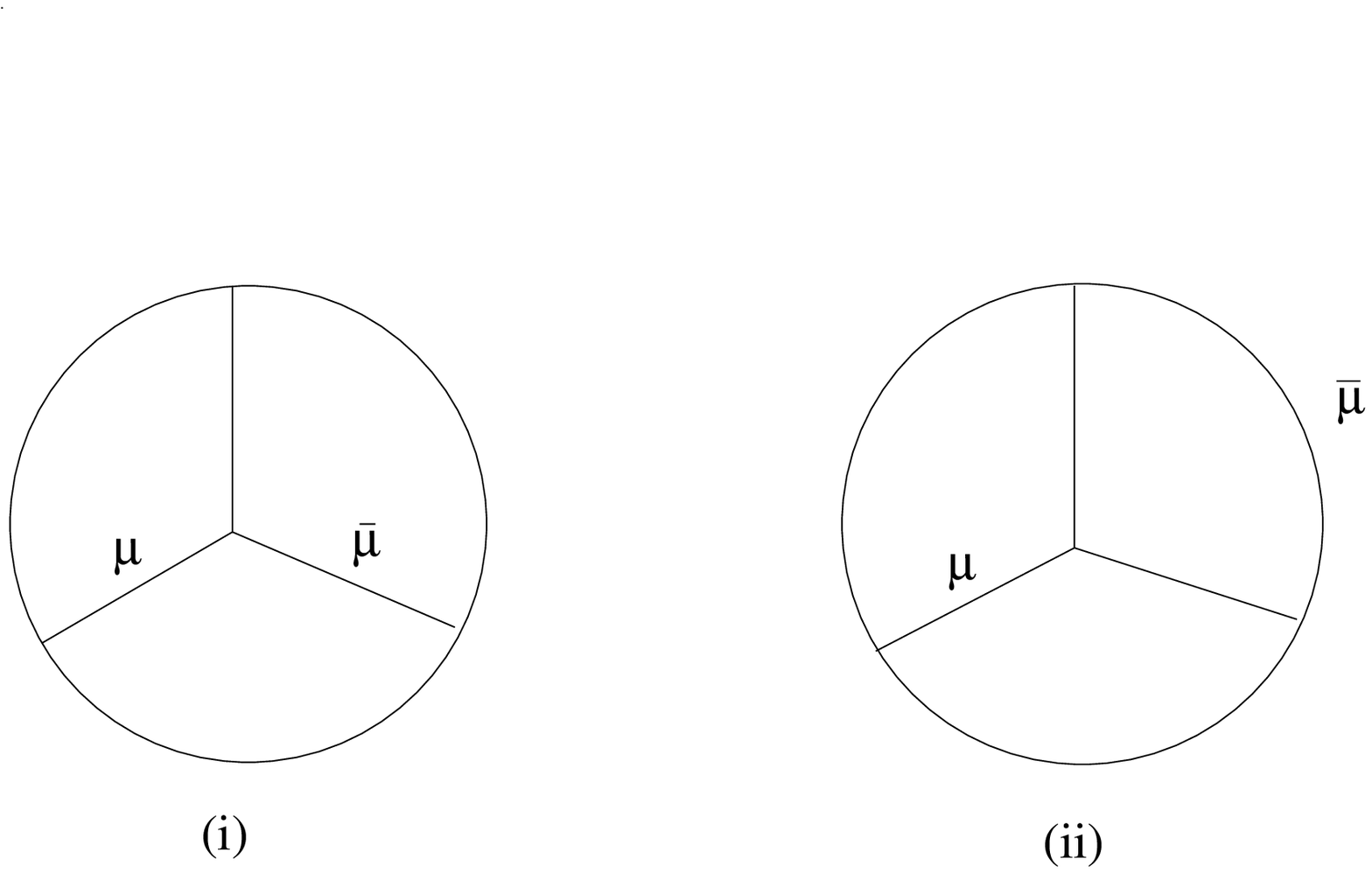}
\end{picture}}
\]
\caption{The diagrams (i) $\mathcal{M}_1$ and (ii) $\mathcal{M}_2$}
\end{center}
\end{figure}

In our analysis, it shall be very convenient to depict the various relations using diagrams. The convention for holomorphic and antiholomorphic derivatives acting on the Green function are given in figure 3.

\begin{figure}[ht]
\begin{center}
\[
\mbox{\begin{picture}(260,50)(0,0)
\includegraphics[scale=.75]{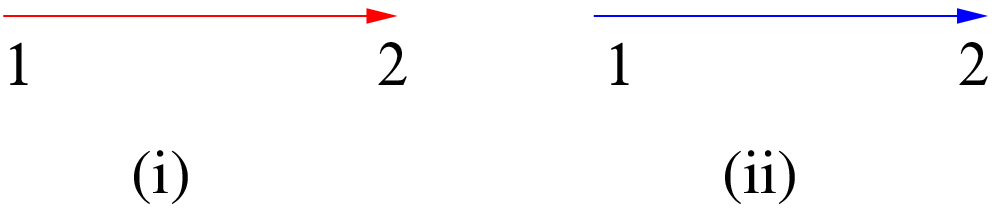}
\end{picture}}
\]
\caption{(i) $\p_2 G_{12} = -\p_1 G_{12}$ (ii) $\bar\p_2 G_{12} = -\bar\p_1 G_{12}$}
\end{center}
\end{figure}

We shall also need the expressions for the various diagrams listed below. One of them is (in the convention of~\cite{D'Hoker:2015foa}) 
\be C_{3,2,1} = \frac{1}{\tau_2^5} \int_{12345} G_{23} G_{34} G_{24} G_{12} G_{16} G_{36} \ee  
as depicted by figure 4. This is a two loop Feynman diagram involving integration over five vertices. Note that this diagram does not involve any derivatives acting on Green functions.

\begin{figure}[ht]
\begin{center}
\[
\mbox{\begin{picture}(110,40)(0,0)
\includegraphics[scale=.35]{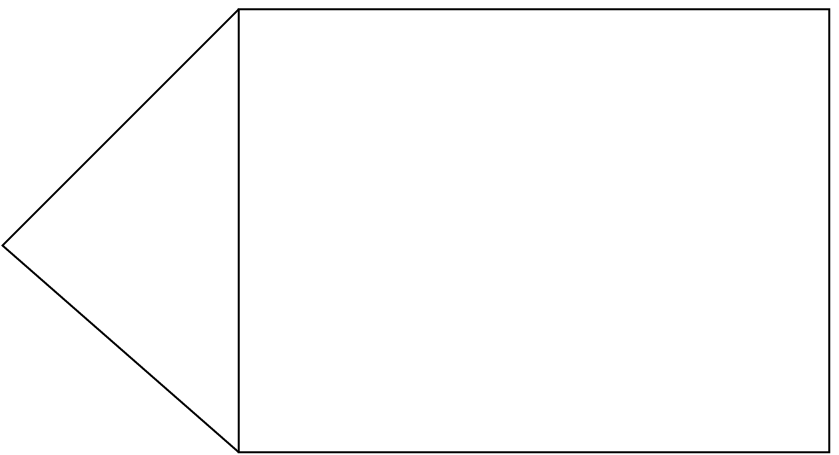}
\end{picture}}
\]
\caption{The diagram $C_{3,2,1}$}
\end{center}
\end{figure}

We also need the diamond diagrams $\mathcal{D}_1, \mathcal{D}_2$ and $\mathcal{D}_3$ in the intermediate steps, defined by
\bea \mathcal{D}_1 = \frac{1}{\tau_2^5} \int_{12345} \p_2 G_{12} \bar\p_1 G_{15} G_{35} G_{23} G_{34} G_{14} G_{45},  \non \\ \mathcal{D}_2 = \frac{1}{\tau_2^5} \int_{12345} \p_1 G_{15} \bar\p_2 G_{23} G_{12} G_{35} G_{14} G_{45} G_{34}, \non \\ \mathcal{D}_3 = \frac{1}{\tau_2^5} \int_{12345} \p_3 G_{35} \bar\p_4 G_{34} G_{45} G_{14} G_{15} G_{12} G_{23}, \eea 
and depicted by figure 5. These are three loop diagrams that involve integrals over five vertices. Each diagram has one holomorphic and one antiholomorphic derivative acting on distinct Green functions.   

\begin{figure}[ht]
\begin{center}
\[
\mbox{\begin{picture}(350,130)(0,0)
\includegraphics[scale=.7]{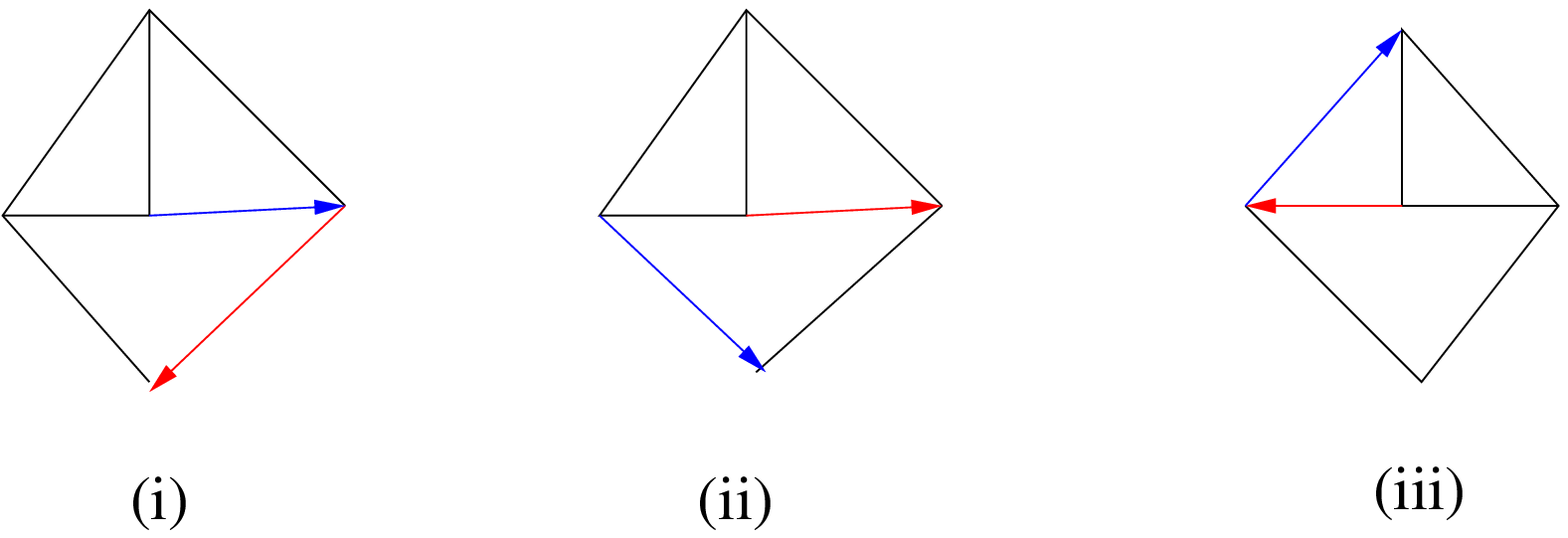}
\end{picture}}
\]
\caption{The diamond diagrams (i) $\mathcal{D}_1$, (ii) $\mathcal{D}_2$, (iii) $\mathcal{D}_3$}
\end{center}
\end{figure}

\begin{figure}[ht]
\begin{center}
\[
\mbox{\begin{picture}(370,90)(0,0)
\includegraphics[scale=.5]{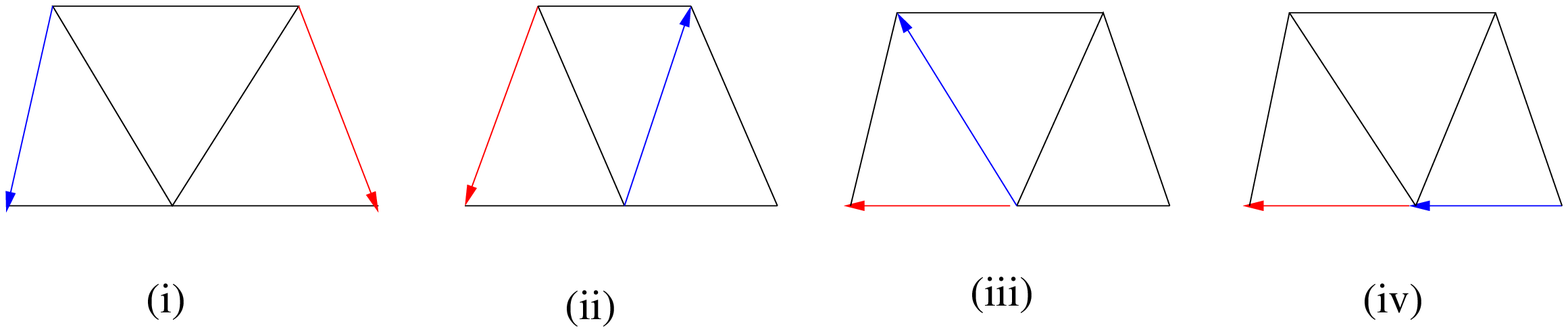}
\end{picture}}
\]
\caption{The fan diagrams (i) $\mathcal{F}_1$, (ii) $\mathcal{F}_2$, (iii) $\mathcal{F}_3$, (iv) $\mathcal{F}_4$}
\end{center}
\end{figure}

Next we list the fan diagrams $\mathcal{F}_1, \mathcal{F}_2, \mathcal{F}_3$ and $\mathcal{F}_4$ which also arise in the intermediate steps, which are defined by
\bea \mathcal{F}_1 = \frac{1}{\tau_2^5} \int_{12345} \p_3 G_{35} \bar\p_1 G_{14} G_{12} G_{23} G_{24} G_{25} G_{45}, \non \\ \mathcal{F}_2 = \frac{1}{\tau_2^5} \int_{12345} \p_1 G_{14} \bar\p_5 G_{25} G_{12} G_{23} G_{24} G_{35} G_{45}, \non \\ \mathcal{F}_3 = \frac{1}{\tau_2^5} \int_{12345} \p_1 G_{12} \bar\p_4 G_{24} G_{23} G_{25} G_{35} G_{14}G_{45},\non \\ \mathcal{F}_4 = \frac{1}{\tau_2^5} \int_{12345}\p_1 G_{12} \bar\p_2 G_{23}G_{14} G_{24} G_{25}G_{35}G_{45} \eea 
and depicted by figure 6. Again, these are three loop diagrams that involve integrals over five vertices. Each diagram has one holomorphic and one antiholomorphic derivative acting on distinct Green functions.

\begin{figure}[ht]
\begin{center}
\[
\mbox{\begin{picture}(310,100)(0,0)
\includegraphics[scale=.6]{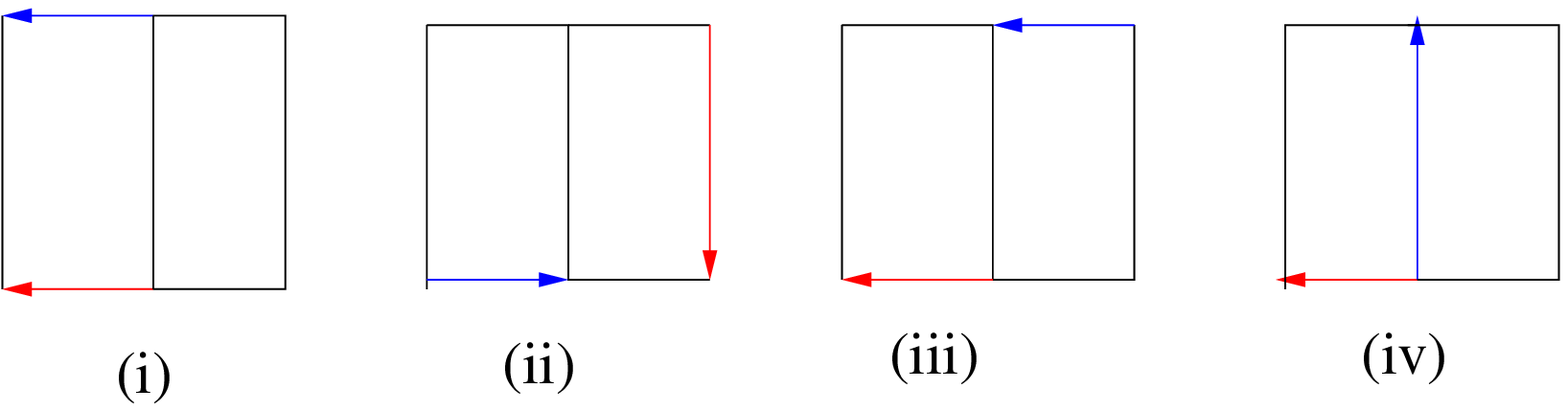}
\end{picture}}
\]
\caption{The ladder diagrams (i) $\mathcal{L}_1$, (ii) $\mathcal{L}_2$, (iii) $\mathcal{L}_3$, (iv) $\mathcal{L}_4$}
\end{center}
\end{figure}

Finally, we consider the ladder diagrams $\mathcal{L}_1, \mathcal{L}_2, \mathcal{L}_3$ and $\mathcal{L}_4$ as they are needed in the intermediate steps, which are defined by
\bea \mathcal{L}_1 =\frac{1}{\tau_2^6}\int_{123456} \p_2 G_{24} \bar\p_1 G_{13} G_{12} G_{34} G_{35} G_{46} G_{56} ,\non \\  \mathcal{L}_2 = \frac{1}{\tau_2^6}\int_{123456}\p_6 G_{56} \bar\p_4 G_{24} G_{12} G_{34} G_{13} G_{35} G_{46},\non \\  \mathcal{L}_3 = \frac{1}{\tau_2^6}\int_{123456} \p_2 G_{24} \bar\p_3 G_{35} G_{12} G_{34} G_{56}G_{13} G_{46} ,\non \\ \mathcal{L}_4 = \frac{1}{\tau_2^6}\int_{123456}\p_2 G_{24} \bar\p_3 G_{34} G_{12} G_{13} G_{35} G_{46} G_{56}.\eea
and depicted by figure 7. These are two loop diagrams that involve integrals over six vertices. Each diagram has one holomorphic and one antiholomorphic derivative acting on distinct Green functions.

Now let us first consider the contribution coming from $\mathcal{M}_1$ in \C{M12}. We have that
\be \mathcal{M}_1 = -\frac{\tau_2}{\pi} (\mathcal{D}_1 +\mathcal{D}_2)+C_{3,2,1}-\frac{\tau_2}{\pi} \mathcal{F}_1.\ee
Using the identity\footnote{It is easy to see that $\mathcal{D}_1$ remains the same when $\p$ and $\bar\p$ are interchanged in the integrand.} 
\be \frac{\tau_2}{\pi} \mathcal{D}_1 = -\frac{\tau_2}{\pi} \mathcal{D}_2= -\frac{\mathcal{M}}{2}\ee
depicted by figure 8, this gives us
\be \label{m1}\mathcal{M}_1 = C_{3,2,1} -\frac{\tau_2}{\pi} \mathcal{F}_1,\ee 
as the sum of the two diamond diagrams cancel.

\begin{figure}[ht]
\begin{center}
\[
\mbox{\begin{picture}(380,80)(0,0)
\includegraphics[scale=.55]{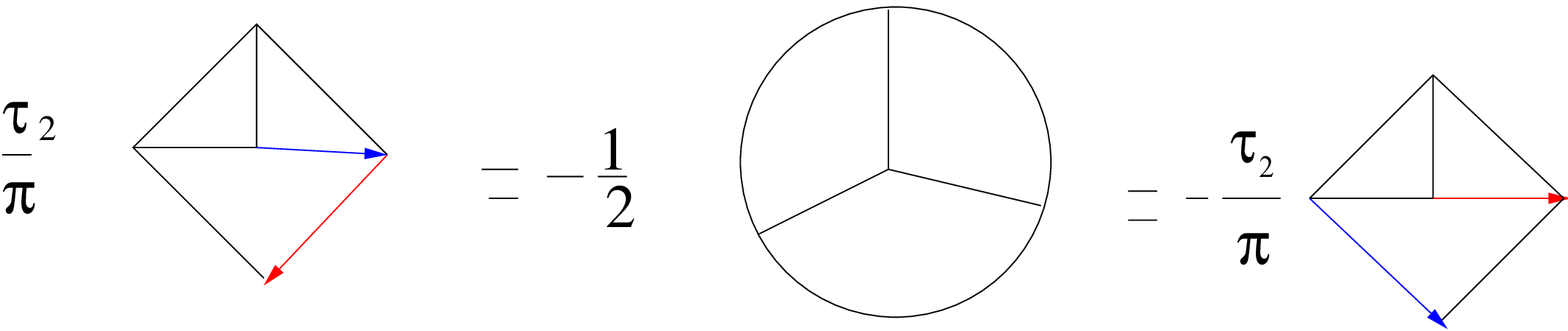}
\end{picture}}
\]
\caption{The relation $\tau_2\mathcal{D}_1/\pi = - \tau_2 \mathcal{D}_2/\pi= -\mathcal{M}/2$}
\end{center}
\end{figure}

Next let us consider the contribution from $\mathcal{M}_2$. We have that
\be \label{C2}\frac{\mathcal{M}_2}{2} = \frac{\tau_2}{\pi}\Big(\mathcal{L}_1 +\mathcal{L}_2 +\mathcal{L}_3 +\mathcal{F}_1 +\mathcal{F}_2 -\mathcal{F}_3 - \mathcal{D}_3\Big) . \ee
On using the identities for the ladder diagrams 
\bea \frac{\tau_2}{\pi}\mathcal{L}_1 = C_{3,2,1}, \quad \mathcal{L}_2 = \mathcal{L}_3\eea
depicted by figures 9 and 10 respectively,
we get that
\bea  \frac{\tau_2}{\pi}\Big(\mathcal{L}_1 +\mathcal{L}_2 +\mathcal{L}_3 \Big) = C_{3,2,1} +2\mathcal{L}_2 = C_{3,2,1} +2\mathcal{L}_3.\eea

\begin{figure}[ht]
\begin{center}
\[
\mbox{\begin{picture}(310,60)(0,0)
\includegraphics[scale=.6]{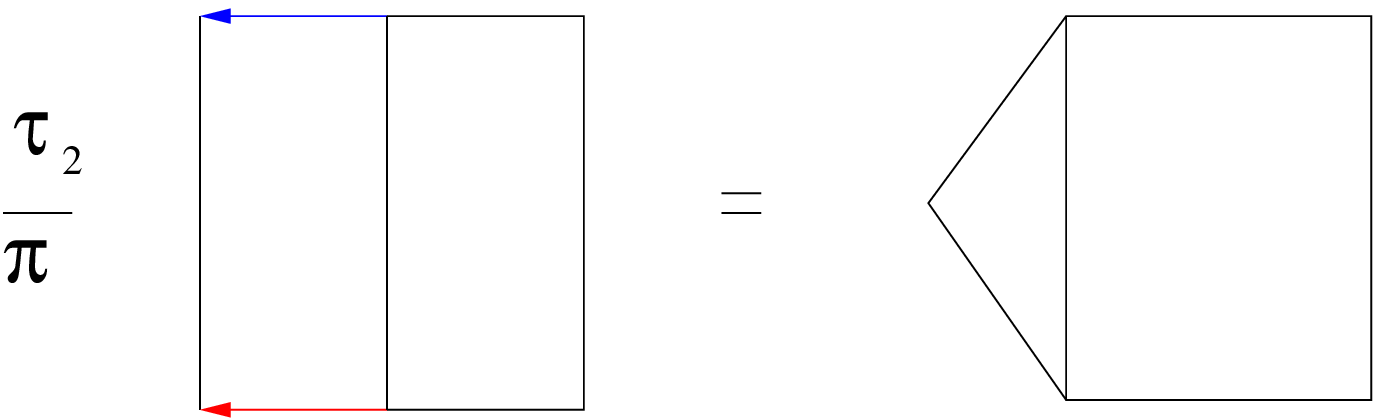}
\end{picture}}
\]
\caption{The relation $\tau_2\mathcal{L}_1/\pi = C_{3,2,1}$}
\end{center}
\end{figure}

\begin{figure}[ht]
\begin{center}
\[
\mbox{\begin{picture}(220,40)(0,0)
\includegraphics[scale=.6]{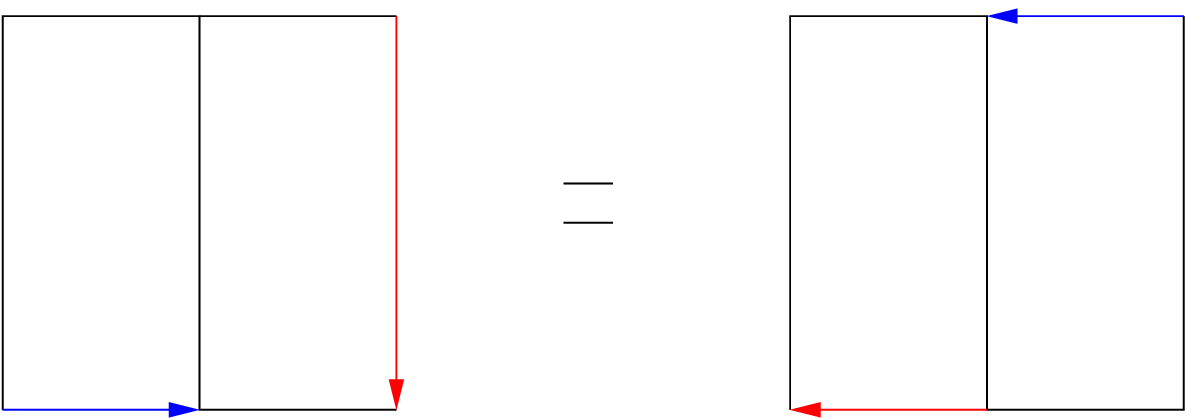}
\end{picture}}
\]
\caption{The relation $\mathcal{L}_2 = \mathcal{L}_3$}
\end{center}
\end{figure}

Now to solve for $\mathcal{L}_3$ we note that
\be \frac{\tau_2}{\pi} \mathcal{L}_3 = \frac{\tau_2}{\pi}(\mathcal{L}_1-\mathcal{L}_4)= C_{3,2,1} - \frac{\tau_2}{\pi}\mathcal{L}_4. \ee
We directly solve for $\mathcal{L}_4$ to obtain
\be \frac{2\tau_2}{\pi}\mathcal{L}_4 = E_3^2- E_6,\ee
as depicted by figure 11. This leads to
\be \frac{\tau_2}{\pi}\mathcal{L}_3 = C_{3,2,1} - \frac{E_3^2}{2} +\frac{E_6}{2},\ee
as depicted by figure 12. 
Thus we finally get that
\be  \frac{\tau_2}{\pi}\Big(\mathcal{L}_1 +\mathcal{L}_2 +\mathcal{L}_3 \Big)  = 3 C_{3,2,1} - E_3^2 +E_6.\ee

\begin{figure}[ht]
\begin{center}
\[
\mbox{\begin{picture}(380,100)(0,0)
\includegraphics[scale=.6]{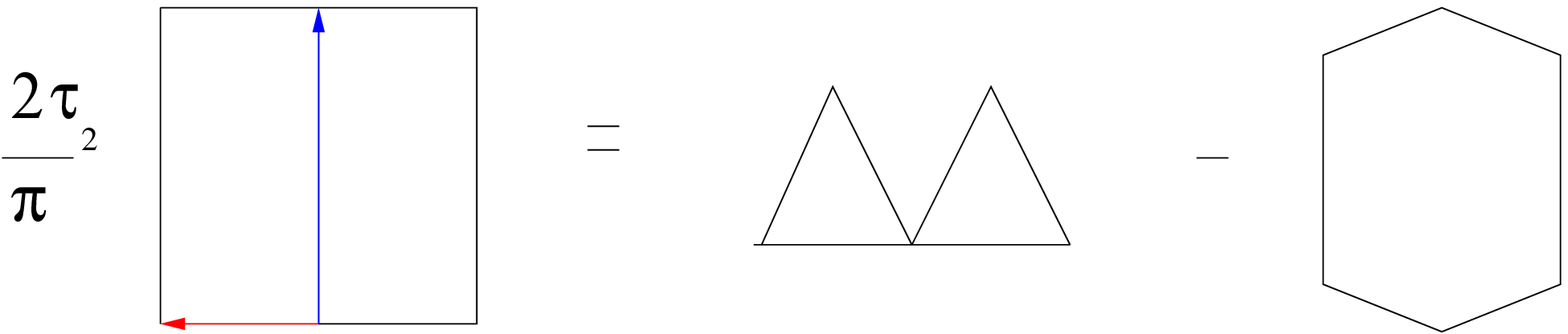}
\end{picture}}
\]
\caption{The relation $2\tau_2 \mathcal{L}_4/\pi = E_3^2 - E_6$}
\end{center}
\end{figure}

\begin{figure}[ht]
\begin{center}
\[
\mbox{\begin{picture}(380,40)(0,0)
\includegraphics[scale=.5]{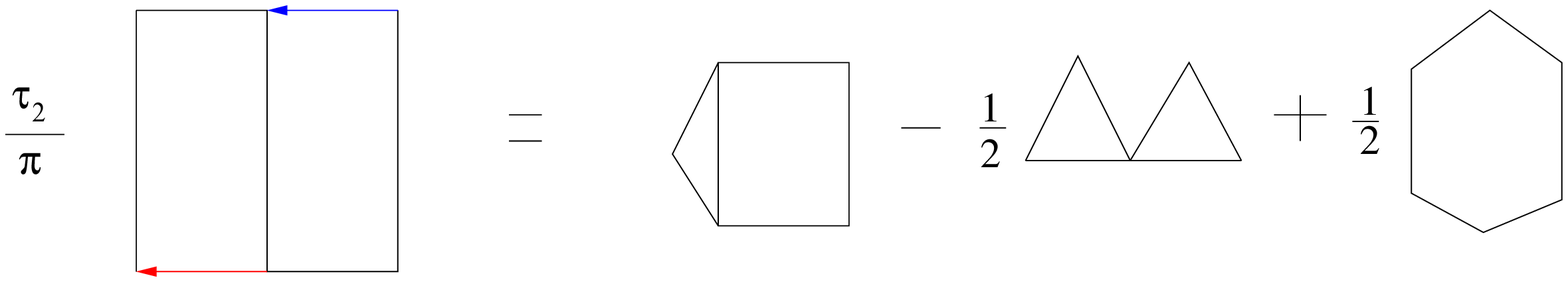}
\end{picture}}
\]
\caption{The relation $\tau_2 \mathcal{L}_3/\pi = C_{3,2,1}-E_3^2/2 + E_6/2$}
\end{center}
\end{figure}

Let us consider the remaining contributions to $\mathcal{M}_2/2$ in \C{C2}.
We have that
\be \frac{\tau_2}{\pi} (\mathcal{F}_2 -\mathcal{F}_3)= -D_{2,1,1,1;1}-\frac{\tau_2}{\pi}\mathcal{F}_4, \ee
as depicted by figure 13. Here we have a new diagram $D_{2,1,1,1;1}$ defined by
\be D_{2,1,1,1;1}= \frac{1}{\tau_2^4} \int_{1234} G_{13} G_{23} G_{12} G_{24} G_{14}^2\ee
as depicted by figure 14. This diagram also arises in the expression for the $D^{12} \mathcal{R}^4$ amplitude at genus one. Finally we use the relation
\be \mathcal{F}_4 = -\mathcal{F}_1\ee
as depicted by figure 15, to obtain
\be \frac{\tau_2}{\pi}\Big(\mathcal{F}_1 +\mathcal{F}_2 -\mathcal{F}_3 - \mathcal{D}_3 \Big) = -D_{2,1,1,1;1} +\frac{2\tau_2}{\pi} \mathcal{F}_1 - \frac{\tau_2}{\pi} \mathcal{D}_3.\ee

\begin{figure}[ht]
\begin{center}
\[
\mbox{\begin{picture}(270,50)(70,0)
\includegraphics[scale=.55]{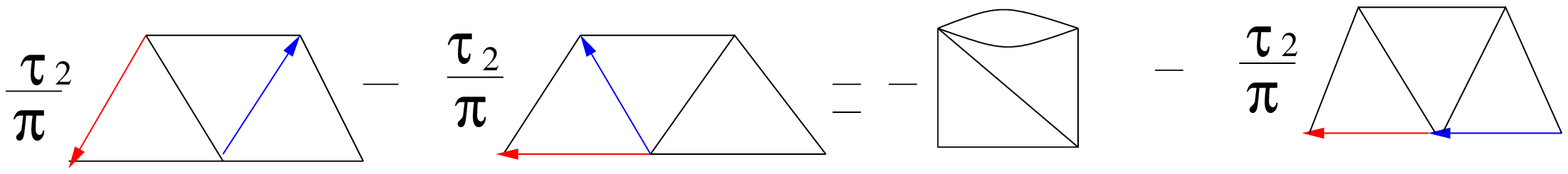}
\end{picture}}
\]
\caption{The relation $\tau_2\mathcal{F}_2/\pi - \tau_2\mathcal{F}_3/\pi = -D_{2,1,1,1;1}- \tau_2\mathcal{F}_4/\pi$}
\end{center}
\end{figure}

\begin{figure}[ht]
\begin{center}
\[
\mbox{\begin{picture}(100,70)(0,0)
\includegraphics[scale=.55]{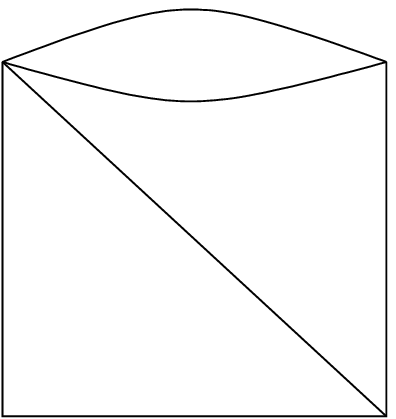}
\end{picture}}
\]
\caption{The diagram $D_{2,1,1,1;1}$}
\end{center}
\end{figure}

\begin{figure}[ht]
\begin{center}
\[
\mbox{\begin{picture}(160,40)(70,0)
\includegraphics[scale=.75]{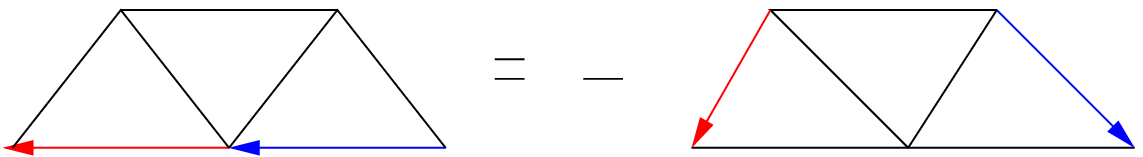}
\end{picture}}
\]
\caption{The relation $\mathcal{F}_4  = - \mathcal{F}_1$}
\end{center}
\end{figure}

\begin{figure}[ht]
\begin{center}
\[
\mbox{\begin{picture}(160,50)(70,0)
\includegraphics[scale=.6]{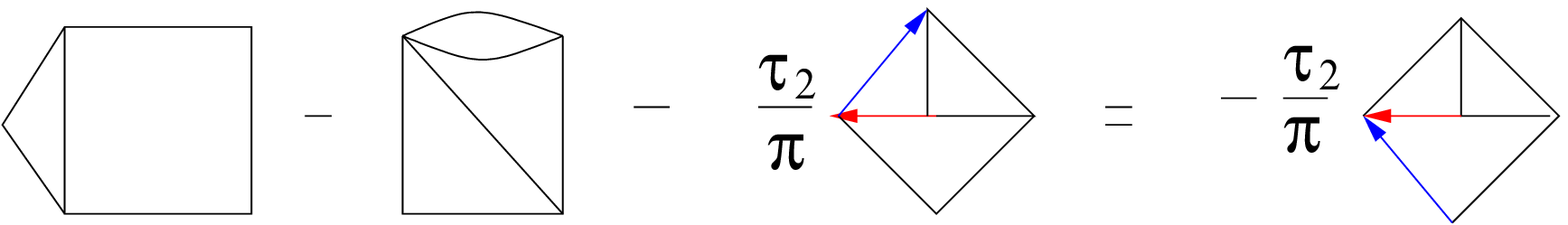}
\end{picture}}
\]
\caption{The relation $C_{3,2,1} - D_{2,1,1,1;1} - \tau_2 \mathcal{D}_3/\pi = \tau_2{\mathcal{D}}_1/\pi = -\mathcal{M}/2$}
\end{center}
\end{figure}

Thus putting all the contributions together, we get that
\be \label{m2}\frac{\mathcal{M}_2}{2} = 3 C_{3,2,1} -E_3^2 + E_6 -D_{2,1,1,1;1} +\frac{2\tau_2}{\pi} \mathcal{F}_1 - \frac{\tau_2}{\pi} \mathcal{D}_3.\ee

\begin{figure}[ht]
\begin{center}
\[
\mbox{\begin{picture}(220,50)(90,0)
\includegraphics[scale=.55]{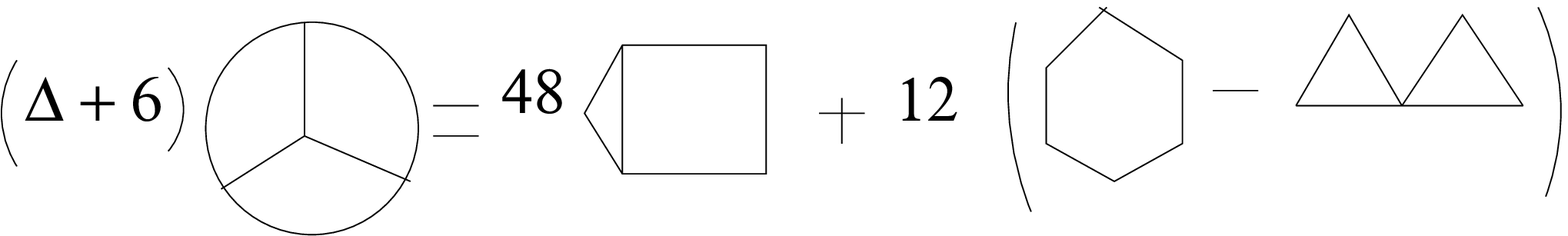}
\end{picture}}
\]
\caption{The Poisson equation for $\mathcal{M}$}
\end{center}
\end{figure}

Now the expression \C{m2} simplifies further by using the relation
\be C_{3,2,1} - D_{2,1,1,1;1} -\frac{\tau_2}{\pi}\mathcal{D}_3 = \frac{\tau_2}{\pi}{\mathcal{D}}_1 = -\frac{\mathcal{M}}{2}\ee 
as depicted by figure 16, leading to
\be \label{m3}\frac{\mathcal{M}_2}{2} = 2C_{3,2,1} -E_3^2 + E_6 -\frac{\mathcal{M}}{2} +\frac{2\tau_2}{\pi}\mathcal{F}_1.\ee

Now from \C{m1} and \C{m3} we see that the total contribution involving $\mathcal{F}_1$ exactly cancels in $24\mathcal{M}_1 +6\mathcal{M}_2$, resulting in considerable simplification. 

This leads to the Poisson equation 
\be \label{p}(\Delta +6)\mathcal{M} = 48 C_{3,2,1} + 12 (E_6 - E_3^2)\ee
satisfied by the Mercedes diagram, as depicted in figure 17. Thus the source terms involve simple one and two loop Feynman diagrams.

\subsection{An elementary consistency check}

We now perform an elementary consistency check of \C{p}. We show that the non--vanishing contributions as $\tau_2 \rightarrow \infty$ match on both sides of the equation. 

In order to obtain these contributions for $C_{3,2,1}$, we note that $C_{3,2,1}$ and $C_{2,2,2}$ satisfy the coupled Poisson equations~\cite{D'Hoker:2015foa}
\bea \label{coupled}(\Delta -8) C_{3,2,1} &=& -C_{2,2,2} - 4 (E_3^2 -4 E_6), \non \\ (\Delta -6) C_{2,2,2} &=& -24 C_{3,2,1} + 12 (E_3^2 - E_6).\eea
Here $C_{2,2,2}$ is a three loop Feynman diagram with five vertices defined by (in the convention of~\cite{D'Hoker:2015foa})
\be C_{2,2,2} = \frac{1}{\tau_2^5} \int_{12345} G_{13} G_{23} G_{14} G_{24} G_{15} G_{25} \ee 
as given in figure 18.

\begin{figure}[ht]
\begin{center}
\[
\mbox{\begin{picture}(80,40)(0,0)
\includegraphics[scale=.65]{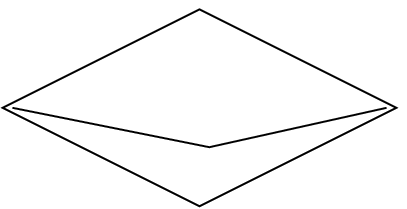}
\end{picture}}
\]
\caption{The diagram $C_{2,2,2}$}
\end{center}
\end{figure}

Thus from \C{coupled} we get that
\bea \label{coupled2}(\Delta -2) (4 C_{3,2,1} + C_{2,2,2}) &=& 52 E_6 - 4 E_3^2, \non \\ 
(\Delta -12) (6 C_{3,2,1} - C_{2,2,2}) &=& 108 E_6 - 36 E_3^2.\eea

Now let us consider the contributions from terms that diverge or are constant as $\tau_2 \rightarrow \infty$, where we make use the expressions
\bea \label{E1}E_6 &=& \frac{2}{\pi^6} \zeta(12) \tau_2^6+\ldots, \non \\ E_3^2 &=& \frac{4}{\pi^6}\zeta(6)^2 \tau_2^6 +\frac{3}{\pi^5} \zeta(5)\zeta(6)\tau_2+\ldots,\eea
where we have ignored terms that vanish as $\tau_2 \rightarrow \infty$, which is also true of the various expressions below. Solving \C{coupled2} we get that
\bea 4 C_{3,2,1} + C_{2,2,2} = \frac{2158}{691\pi^6} \zeta(12) \tau_2^6+c_1 \tau_2^2 + \frac{6}{\pi^5} \zeta(5)\zeta(6)\tau_2+\ldots, \non \\ 6 C_{3,2,1} - C_{2,2,2} = \frac{2572}{691\pi^6} \zeta(12) \tau_2^6 + c_2 \tau_2^4 + \frac{9}{\pi^5} \zeta(5)\zeta(6)\tau_2 +\ldots,\eea
where $c_1$ and $c_2$ are undetermined constants, as they are the zero modes of \C{coupled2}. This leads to the expression for $C_{3,2,1}$ given by
\be \label{E2}10 C_{3,2,1} = \frac{4730}{691\pi^6} \zeta(12) \tau_2^6 + c_2 \tau_2^4 + c_1 \tau_2^2 +\frac{15}{\pi^5} \zeta(5)\zeta(6)\tau_2 +\ldots.\ee
However $c_1$ and $c_2$ are easily seen to vanish as discussed in appendix D.1 of~\cite{Green:2008uj}. We also get that
\be \label{E3}C_{2,2,2} = \frac{266}{691\pi^6} \zeta(12)\tau_2^6+\ldots,\ee
as the $O(\tau_2)$ term cancels.
Thus from \C{E1} and \C{E2} we get that
\be 48 C_{3,2,1} + 12 (E_6 - E_3^2)= \frac{4968}{691\pi^6} \zeta(12)\tau_2^6 + \frac{36}{\pi^5} \zeta(5)\zeta(6)\tau_2 +\ldots\ee
which appears on the right hand side of the Poisson equation \C{p}.
This precisely agrees with $(\Delta+ 6)\mathcal{M}$ which appears on the left hand side of \C{p} on using the large $\tau_2$ expansion~\cite{Green:2008uj}
\be \label{valM}\mathcal{M} = \frac{138}{691\pi^6} \zeta(12)\tau_2^6 + \frac{6}{\pi^5} \zeta(5)\zeta(6)\tau_2\ldots .\ee

\section{The contribution of the Mercedes diagram to the genus one $D^{12} \mathcal{R}^4$ amplitude}

The Mercedes diagram contributes~\cite{Green:2008uj}
\be \label{comp}I^{D^{12}\mathcal{R}^4} = \frac{2\pi \cdot80}{6!} \s_3^2 \mathcal{R}^4 \int_{\mathcal{F}_L} \frac{d^2\tau}{\tau_2^2} \mathcal{M}\ee
to the genus one $D^{12} \mathcal{R}^4$ amplitude which follows from \C{cont}, which we now evaluate. Using \C{p} and the eigenvalue equation \C{eisenstein} for $E_6$, we get that
\be \int_{\mathcal{F}_L} \frac{d^2\tau}{\tau_2^2} \mathcal{M} =  \frac{1}{6}\int_{\mathcal{F}_L} \frac{d^2\tau}{\tau_2^2} \Big(-\Delta \mathcal{M} +\frac{2}{5} \Delta E_6 -12 E_3^2 + 48 C_{3,2,1} \Big).\ee
From \C{coupled2} we also obtain that
\be C_{3,2,1} = \frac{1}{24}\Delta (6 C_{3,2,1} + C_{2,2,2}) - \frac{7}{2} E_6 + \frac{1}{2} E_3^2.\ee
Thus we get that
\bea \label{calc}\int_{\mathcal{F}_L} \frac{d^2\tau}{\tau_2^2} \mathcal{M} &=& \int_{\mathcal{F}_L} \frac{d^2\tau}{\tau_2^2}\Delta \Big(-\frac{\mathcal{M}}{6} + 2 C_{3,2,1} + \frac{1}{3} C_{2,2,2} -\frac{13}{15}E_6\Big) +2\int_{\mathcal{F}_L} \frac{d^2\tau}{\tau_2^2} E_3^2,\eea
on using \C{eisenstein} again for $E_6$. The first integral on the right hand side of \C{calc} is a boundary term which receives contribution only from $\tau_2 \rightarrow \infty$. This evaluates to
\be \int_{-1/2}^{1/2} d\tau_1 \frac{\p}{\p\tau_2}\Big(-\frac{\mathcal{M}}{6} + 2 C_{3,2,1} + \frac{1}{3} C_{2,2,2} -\frac{13}{15}E_6\Big)\Big\vert_{\tau_2 =L\rightarrow \infty},\ee
which we now obtain using \C{valM}, \C{E2}, \C{E3} and \C{E1}. Apart from an $O(L^5)$ contribution which must cancel from the integral over $\mathcal{R}_L$, there is a finite contribution equal to
\be \label{1c}\frac{2}{\pi^5} \zeta(5)\zeta(6).\ee
The second integral on the right hand side of \C{calc} can be directly evaluated using~\cite{Zagier2,Green:1999pv,Green:2008uj} 
\be \label{exact}\frac{\pi^{2s}}{4 \zeta (2s)^2}\int_{\mathcal{F}_L} \frac{d^2\tau}{\tau_2^2} E_s^2 = \frac{L^{2s-1}}{2s-1} +2\phi(s) {\rm ln}\Big(\frac{L}{\mu_{2s}}\Big)+\ldots\ee
for $s >1/2$, where we have dropped terms that vanish as $L\rightarrow \infty$. This can be obtained using the explicit expression for the non--holomorphic Eisenstein  series $E_s$. In \C{exact} we have that
\bea \phi(s) &=& \sqrt{\pi} \frac{\Gamma(s-1/2)\zeta(2s-1)}{\Gamma(s)\zeta(2s)}, \non \\ {\rm ln}\mu_{2s} &=& \frac{\zeta'(2s-1)}{\zeta(2s-1)} -\frac{\zeta'(2s)}{\zeta(2s)}+\frac{\Gamma'(s-1/2)}{2\Gamma(s-1/2)} - \frac{\Gamma'(s)}{2\Gamma(s)}.\eea
Thus apart from terms that diverge as $L^5$ and ${\rm ln}L$, we get a finite contribution from the scale of the logarithm. This finite contribution to 
\be \label{log}\int_{\mathcal{F}_L} \frac{d^2\tau}{\tau_2^2} E_3^2 \ee
is equal to
\be \label{2c}-\frac{3}{\pi^5} \zeta(5)\zeta(6) {\rm ln}\mu_6,\ee
where
\be {\rm ln}\mu_6 = \frac{\zeta'(5)}{\zeta(5)} -\frac{\zeta'(6)}{\zeta(6)} +\frac{7}{12} - {\rm ln}2,\ee
on using
\be \frac{\Gamma'(5/2)}{\Gamma(5/2)} = \frac{8}{3} -\gamma -{\rm ln}4, \quad \frac{\Gamma'(3)}{\Gamma(3)} = \frac{3}{2} -\gamma.\ee
Thus adding the contributions \C{1c} and \C{2c}, we get a finite non--vanishing contribution given by
\be \int_{\mathcal{F}_L} \frac{d^2\tau}{\tau_2^2} \mathcal{M}  = -\frac{6\zeta(5)\zeta(6)}{\pi^5} \Big(\frac{\zeta'(5)}{\zeta(5)}-\frac{\zeta'(6)}{\zeta(6)} +\frac{1}{4} -{\rm ln}2\Big),\ee
leading to
\be I^{D^{12}\mathcal{R}^4}=-\frac{4}{3\pi^4} \zeta(5)\zeta(6) \Big(\frac{\zeta'(5)}{\zeta(5)}-\frac{\zeta'(6)}{\zeta(6)} +\frac{1}{4} -{\rm ln}2\Big) \s_3^2\mathcal{R}^4.\ee
In particular, note that there is a term which diverges as ${\rm ln}L$ in \C{log}. This must be cancelled by a term schematically of the form $\s_3^2\mathcal{R}^4{\rm ln}(\alpha' L\mu s)$ coming from the integral over $\mathcal{R}_L$, where $\mu$ is a constant. This leads to a non--analytic term in the external momenta in the effective action.  

\section{Discussion}

We have considered a particularly simple  Feynman diagram which contributes to the $D^{12} \mathcal{R}^4$ amplitude at genus one, and showed that it satisfies a modular invariant Poisson equation, with a very specific structure of source terms. Clearly it would be interesting to generalize the analysis and obtain Poisson equations for the other Feynman diagrams at this order, and at higher orders in the low momentum expansion, which are needed to obtain their contribution to the string amplitude at genus one. Along with a similar analysis for other amplitudes, this will give us a detailed understanding of non--BPS interactions at genus one. Such interactions are not well understood, apart from some analysis based on constraints due to supersymmetry, and multi--loop supergravity~\cite{Green:2008bf,Basu:2008cf,Basu:2013goa,Basu:2013oka}. Also it would be interesting to generalize the analysis to amplitudes at higher genus in string theory. Explicit expressions for the four graviton amplitude at three and four loops in maximal supergravity~\cite{Bern:2008pv,Bern:2009kd,Basu:2014hsa,Basu:2014uba,Basu:2015dsa} might provide useful hints about the structure of the corresponding string amplitudes.        

\providecommand{\href}[2]{#2}\begingroup\raggedright\endgroup

\end{document}